\documentstyle[preprint,aps,eqsecnum,epsfig,prb]{revtex}

\draft

\tighten

\title{Anisotropic scattering and quantum 
magnetoresistivities of a periodically modulated 2D electron gas}

\author{Andrei Manolescu 
 }

\address{
Institutul Na\c{t}ional de Fizica Materialelor, C.P. MG-7 
Bucure\c{s}ti-M\u{a}gurele,
  Romania} 

\author{Rolf R. Gerhardts}
\address{Max-Planck-Institut f\"ur Festk\"orperforschung, Heisenbergstra\ss e
  1,
D-70569 Stuttgart, Germany}
\author{Michael Suhrke and Ulrich R\"ossler} 

\address{Institute for Theoretical Physics, University of Regensburg, D-93040
  Regensburg, Germany}

\begin{document}

\maketitle

\begin{abstract}
We calculate the longitudinal conductivities of a two-dimensional
noninteracting electron gas in a uniform magnetic field and a lateral
electric or magnetic periodic modulation in one spatial direction, in
the quantum regime.  We consider the effects of the electron-impurity
scattering anisotropy through the vertex corrections on the Kubo
formula, which are calculated with the Bethe-Salpeter equation,
in the self-consistent Born approximation.  We find that due to the
scattering anisotropy the band conductivity increases, and the scattering
conductivities decrease and become anisotropic.  Our results are in 
qualitative agreement with recent experiments.
\end{abstract}


\section{Introduction}

Magnetotransport properties of  two-dimensional electron gases (2DEGs)
at semiconductor interfaces,
subjected to one-dimensional lateral superlattices defined by periodic
electrostatic or magnetostatic fields, have 
%
attracted increasing interest
during the last decade.  Properties like the commensurability
(Weiss) oscillations of the magnetoresistance \cite{Weiss89:179} or
the low-field positive magnetoresistance \cite{Beton90:9229} are well
understood and have been qualitatively explained both by quantum-mechanical
\cite{Gerhardts89:1173,Zhang90:12850,Xue92:5986,Shi96:12990}
and by classical transport calculations.
\cite{Beenakker89:2020,Gerhardts96:11064}  
These interesting effects occur at low temperatures (about 4~K or below), where
the transport properties are governed by random-impurity scattering.
 It is well known that in the high-mobility 2DEGs at
semiconductor interfaces the basic scattering
mechanism is due to long-range Coulomb impurities, which lead predominantly to
small-angle scattering of the electrons.
 In all the mentioned papers this aspect has, however, 
been neglected and the electron-impurity scattering
has been treated within simplified, phenomenological models, related to
a simple relaxation-time approximation.  Such models are suitable only for
short-range (delta-potential) impurities, which lead to isotropic 
scattering, in contrast to the anisotropic scattering implied by 
realistic impurities.

Only the most recent calculations, based on classical mechanics
and Boltzmann equation, have proven that an adequate treatment of scattering
anisotropy 
is important for an understanding of the experimental data.  For
instance, the number and the amplitude of the resolved
Weiss oscillations of the resistivity component $\rho_{\perp}$, measured when
the current 
flows  {\em perpendicular} to the superlattice (i.e.,  perpendicular to the
direction of translational invariance), can be fitted only if a strongly
anisotropic scattering is assumed.\cite{Menne98:1707,Mirlin98:12986} 
Although in  the regime of  Weiss oscillations  the magnetic fields
are apparently weak enough to allow for classical calculations, quantum
effects   determined by the Landau quantization  also  play a role in
this regime.
The (weaker) oscillations of the resistivity component $\rho_{\parallel}$,
which are observed  when the current flows {\em parallel} to the superlattice
and have a phase opposite to those of $\rho_{\perp}$, have been explained by
the quantum oscillations in the density of states (DOS).
\cite{Weiss89:179,Zhang90:12850}  For stronger magnetic fields, the
presence of well defined Landau bands has been clearly identified also
in the perpendicular resistance.\cite{Weiss90:88,Tornow96:16397} More
recently, periodic magnetic fields with extremely strong gradients, of
amplitudes up to 0.4~T and periods of 500~nm, have been obtained, and
huge magnetoresistance oscillations have been detected,\cite{Ye98:330}
presumably of a quantum origin.  Experiments on another class of
unidirectional (electrostatic) superlattices, of short-periods, have shown
an anisotropy of the magnetoresistivity tensor which could be explained, 
within a semi-classical theory, by
different transport times in the directions perpendicular and parallel
to the superlattice, and ascribed to the anisotropic character of the
scattering events. \cite{Petit}


Previous quantum-mechanical calculations for modulated systems
 have treated the electron-impurity self-energy as a simple
c-number.\cite{Zhang90:12850,Tornow96:16397,Steffens98:3859}
Strictly speaking, this is correct only for delta impurities in the unmodulated
system, whereas for modulated systems even in the simple  self-consistent Born
approximation  and for   delta impurities a complicated self-energy
operator results, which does neither commute with the Hamiltonian of the
impurity-free  modulated system, nor with the Green function operator of the
impurity averaged system. The assumption of a c-number self-energy may be
sufficient to explain certain aspects of the influence of a periodic
modulation on the magnetoresistivities  qualitatively, 
especially for a weak modulation.  For a strong modulation however, it can not
 be justified and will lead to  incorrect results. 
 Moreover, to include the effect of strongly
anisotropic impurity scattering, which turned out to be important in the
classical calculations, we have to consider  long-range impurity potentials,
which lead to 
important current vertex corrections and are not compatible with a c-number
approximation for the self-energy.

In this paper we take the mentioned
experimental information as a motivation to perform a quantum-mechanical
transport calculation with a more elaborated treatment of the
electron-impurity scattering, including the scattering anisotropy.
We shall consider only  modulations of the 2DEG varying along
one lateral direction, 
 partly because this
is a situation of considerable experimental relevance, but also since we
expect in this situation 
especially strong anisotropy effects resulting from the interplay of
anisotropic periodic modulation and anisotropic impurity scattering.    
We use the Kubo formalism, and we calculate the 
electron-impurity self-energy in the self-consistent Born approximation
(SCBA), on the same footing with the vertex function, with a numerical
scheme based on Fourier expansions.  For technical reasons we describe
the finite-range impurities by Gaussian potentials.  
After recalling the simplest and most important classical results (Sec.~II and
Appendix B), we describe our calculations (Sec.~III and Appendix A), and then
we  discuss examples with electric and magnetic modulations (Sec.~IV).
Finally we  close with some general conclusions (Sec.~IV).

\section{Simple classical results} \label{classresults}

We sketch briefly the simplest results concerning the effect of the
scattering anisotropy in transport.  For the homogeneous, unmodulated
system, the best known expressions for the conductivities of the 2DEG
in a magnetic field are the classical Drude formulas,
\begin{equation}
\sigma_{xx}=\sigma_{yy}=\frac{\sigma_0}{1+(\omega_c\tau)^2}\,,\quad
-\sigma_{xy}=\sigma_{yx}=
\omega_c\tau\sigma_{xx}\,,
\label{drude}
\end{equation}
$\sigma_0=ne^2\tau/m$ being the zero-field conductivity, and $\omega_c =eB_0/m$
the cyclotron frequency in the externally applied perpendicular magnetic field
$B_0$. $n$ is the electron density, $\tau$ is the relaxation time, and $m$ is
the conduction-band effective mass.
For $\omega_c\tau \gg 1$ the diagonal conductivities are proportional to
the scattering rate, $\sigma_{xx}, \sigma_{yy} \sim 1/\tau$, resulting
from transitions of electrons between closed cyclotron orbits, mediated by
electron-impurity scattering. Those contributions to the conductivity will be
called scattering contributions in the following.
In the presence of a (weak) modulation in $x$ direction, the
guiding centers of the cyclotron orbits  perform a drift motion  in $y$
direction. This drift  leads to an additional contribution to
$\sigma_{yy}$, which in the simplest approximation \cite{Beenakker89:2020}
can be written as
average of the square of the drift velocities taken over all drifting orbits
at the Fermi energy $E_F$,
\begin{equation}
\Delta \sigma_{yy} = \frac{e^2 n \tau}{E_F} \left\langle \bar{v}_y^2
\right\rangle \, . \label{sigma_drift}
\end{equation}
Being due to open orbits with a finite velocity,
the modulation induced contribution $\Delta \sigma_{yy}$ is, similar to
$\sigma_0$, proportional to the scattering time $\tau$ itself and not to
its inverse as  $\sigma_{xx}$. These different $\tau$ dependences lead us to
expect peculiar anisotropies of the conductivity tensor in the presence of
anisotropic impurity scattering.

 For isotropic
electron-impurity scattering, the relaxation time in Eq.~(\ref{drude}) is the
average flight time of an electron  between two scattering events,
$\tau=\tau_{sc}$.  For anisotropic scattering this has to be replaced by the 
transport or momentum relaxation time, $\tau=\tau_{tr}$, 
which is given by
\begin{equation} \label{tautrans}
\frac{1}{\tau_{tr}} = \frac{1}{\tau_{sc}} \,
\int_{-\pi}^\pi \frac{d\theta}{2 \pi} \,w(\theta;k_F)\, (1-\cos\theta) \,,
\end{equation}
where  $w$ is the scattering amplitude for elastic scattering at the Fermi
edge from an initial state ${\bf k_i}$ to a final state ${\bf k_f}$, which
depends only on $|{\bf k_f}|=|{\bf k_i}| =k_F$  and the angle
 $\theta$  between ${\bf k_f}$ and ${\bf k_i}$. For isotropic scattering,
 $w(\theta;k_F) \equiv 1$, and  $\tau_{tr}=\tau_{sc}$.
 In general, however, 
$\tau_{tr} /\tau_{sc} >1$, and this ratio increases with increasing
predominance of small-angle scattering. This means that with increasing
importance of forward scattering, and for $\omega_c \tau \gg 1$, the
conductivity component $\sigma_{xx}$ and the scattering contribution to
$\sigma_{yy}$ should become smaller,\cite{Ando82:437}  whereas $\Delta
\sigma_{yy}$ is expected to become larger, similar to  $\sigma_0$.

Corresponding anisotropies are expected for the  resistivity tensor
\begin{equation}
\rho_{xx(yy,xy)}=\frac{\sigma_{yy(xx,yx)}}
{\sigma_{xx}\sigma_{yy}+\sigma_{xy}^2} \,.
\label{rhotensor}
\end{equation}
For the homogeneous system, the classical Drude resistivity tensor has diagonal
components which are independent of the magnetic field,
$\rho_{xx}=\rho_{yy}=1/\sigma_0$, and thus always {\em decrease} with
increasing importance of forward scattering.
A simple argument for this result is that the dominant contribution to the
electrical resistance comes from the electron backscattering while the
forward scattering gives no contribution.  Obviously, for finite-range
impurities the scattering anisotropy favorizes the forward-scattering
events. For a system with periodic modulation in $x$ direction, the classical
calculations \cite{Beenakker89:2020,Menne98:1707}
 yield a modulation-induced contribution  $\Delta \rho_{xx}$ which, in
 contrast to $1/\sigma_0$, {\em increases} with  increasing  forward
 scattering. 


It has also been emphasized in Refs.~\cite{Beenakker89:2020,Menne98:1707}
that, even for isotropic scattering, the backscattering term in Boltzmann's
equation does not vanish as in homogeneous systems, but rather is important to
guarantee particle conservation, i.e., the equation of continuity. 
This backscattering term, which is the
classical analog of the vertex corrections in the quantum treatment, is
necessary to obtain the correct (quasi-local) $B_0$ dependence of the
modulation-induced resistance correction for large magnetic fields, which is
$\Delta \rho _{xx}/ \rho_0 \sim (B_0)^2$ for electric and  $\Delta
\rho _{xx}/ \rho_0 \sim (B_0)^0$ for pure magnetic modulation.
\cite{Menne98:1707} The simplistic approximation of Eq.~(\ref{sigma_drift})
is not in accord with the equation of continuity and yields asymptotic results
which are by a factor proportional to $(B_0)^{-2}$ too
small.\cite{Menne98:1707} These deficiencies become important if $B_0$ becomes
so large that the cyclotron diameter $2R_c$ of electrons at the Fermi level
becomes smaller than the modulation period $a$. \cite{Beenakker89:2020,Menne98:1707}

\section{Kubo formula and self-consistent equations}

The Hamiltonian of an electron situated in the $(x,y)$ plane, in the
presence of modulating magnetic and electric fields is
\begin{equation}
H=\frac{1}{2m}\left[{\bf p} + e {\bf A}(x)\right]^2 + V(x)\,.
\label{hamiltonian}
\end{equation}
The electric and magnetic modulations are defined by a periodic potential,
$V(x)=V(x+a)$, and by a magnetic field with periodic $z$ component
$B(x)=B(x+a)$, respectively. 
We always assume a non-vanishing average value $B_0\neq 0$ of $B(x)$,
and we shall denote by ${\ell}=(\hbar/eB_0)^{1/2}$
the corresponding magnetic length. For the vector potential we use
the Landau gauge, ${\bf A}(x)=(0,\int_0^x B(x')
dx')$.  In absence of the modulation the eigenstates are the well-known
Landau states $|nX_0\rangle$ [see Eq.~(\ref{landaub})] with
center coordinates $X_0=-\ell ^2 p_y / \hbar$, where $p_y$ is the conserved
canonical momentum.
The modulation lifts the degeneracy with respect to $X_0$, which however
remains a good quantum number due to the translational invariance in
the $y$ direction.  The energy spectrum becomes structured in periodic
bands, $E_{n,X_0}= E_{n,X_0+a}$, with the corresponding modified eigenstates
$|nX_0)$. 
Such modulation-induced bands have recently been calculated for arbitrarily
strong pure and mixed electric and magnetic modulations, and the energy spectra
and eigenstates have been related to the different types of corresponding
classical orbits. \cite{Zwerschke99:5536} 
In the following, we will have  to distinguish between  two relevant basis
sets of the Hilbert space, the {\em Landau basis}  $\{|nX_0\rangle\}$ of
eigenstates of the homogeneous 2DEG, 
and the  {\em modulated basis}  $\{|nX_0)\}$ of eigenstates of the
modulated 2DEG.

We apply the standard procedure of averaging the retarded (advanced) Green's
functions over all the configurations of randomly distributed impurities,
$G^{\pm}\equiv \langle \hat G^{\pm}\rangle _{imp}$, where $(\hat G^{\pm})^{-1}
(E) = 
E - H - V_{imp} \pm i0^+$ and  $ V_{imp}$ is the potential describing a given
impurity configuration. This leads, in the simplest consistent approximation,
to the coupled equations 
\begin{eqnarray}
&&( G^{\pm})^{-1} (E) = E - H -\Sigma^{\pm}(E) \,,  
\label{greenf}\\
&&\Sigma^{\pm}(E)=n_i \int d{\bf R} 
\, u({\bf r}-{\bf R}) \, G^{\pm}(E) \, u({\bf r}-{\bf R}) \,.
\label{scba}\\
\nonumber
\end{eqnarray}
Equation (\ref{greenf}) defines the self-energy operator
$\Sigma^{\pm}(E)=\Delta(E) \mp i \Gamma(E)/2$. The spectral operator 
$A(E)=[G^-(E)-G^+(E)]/(2\pi i)$ and $ \Gamma(E)$ are positive 
(semi-)definite Hermitean operators.  The DOS is
$D(E)=D_0\hbar\omega_c/(\pi a) \sum_n \int_0^a dX_0 (nX_0|G^-(E)|nX_0)$,
where $D_0=m/(\pi\hbar^2)$.  Equation (\ref{scba}) is the SCBA, the 
self-consistent approximation of the lowest order in the impurity 
concentration $n_i$, $u$ being the electron-impurity potential.

The diagonal conductivities are calculated from the Kubo formula,
\cite{Ando82:437,Gerhardts71:126,Gerhardts75:327}
\begin{eqnarray}
&&\sigma_{\alpha\alpha}=\int dE \left[-\frac{d{\cal F}(E)}{dE}\right]
\sigma_{\alpha\alpha}(E) \,, \nonumber\\
&&\sigma_{\alpha\alpha}(E)=\frac{e^2\hbar}{2\pi}\frac{1}{L_x L_y}
{\mathrm Tr} \left\{v_{\alpha}
\left[2F_{\alpha}^{+-}(E)-F_{\alpha}^{++}(E)-F_{\alpha}^{--}(E) 
\right]\right\}\,,
\label{kubof}
\end{eqnarray}
where ${\alpha}=x,y$, \,
$L_x$ and $L_y$ are the linear dimensions of the 2DEG, ${\cal F}$
is the Fermi function, $v_{\alpha}$ is the velocity operator, and 
spin-degeneracy is assumed. The vertex functions
$F_{\alpha}^{\sigma\sigma'}
\equiv \langle \hat G^{\sigma} v_{\alpha} \hat G^{\sigma'} \rangle $
with $\sigma = \pm$ are given by the equations
\begin{eqnarray}
&&F_{\alpha}^{\sigma\sigma'}(E)=G^{\sigma}(E)\, v_{\alpha}\, G^{\sigma'}(E) 
+ G^{\sigma}(E)\, I [
F_{\alpha}^{\sigma\sigma'} (E) ]\, G^{\sigma'}(E) \,,
\label{vertexf}\\
&&I[ F_{\alpha}^{\sigma\sigma'}(E)] =n_i \int d{\bf R} 
\, u({\bf r}-{\bf R}) \, F_{\alpha}^{\sigma\sigma'}(E)
\, u({\bf r}-{\bf R}) \,,
\label{vertexc}\\
\nonumber
\end{eqnarray}
which are similar to Eqs. (\ref{greenf}) and (\ref{scba}).  
Equation  (\ref{vertexf})  is the
Bethe-Salpeter equation for the vertex function, and Eq. (\ref{vertexc})
 defines the vertex corrections 
$I[ F_{\alpha}^{\sigma\sigma'} (E)]$ in the SCBA. A diagrammatic formulation
of the  SCBA is shown in Fig.\ 1. It is sufficient for calculations of the
conductivity if interference effects such as weak or strong localization are
not important.

As a simple and easily tractable model for the impurity potential, we use a
Gaussian model,  
\begin{equation}
u({\bf r})=\frac{u_0}{\pi r_0^2} e^{-(r/r_0)^2} \,,\quad
\tilde u({\bf q})=u_0 e^{-(qr_0)^2/4} \,,
\end{equation}
where $\tilde u$ is the Fourier transform.  Within the SCBA, the scattering
effect of randomly distributed impurities of density $n_i$ is described by only
two parameters, an energy $\Gamma_0^0 = {n_i} 
u_0^2 \, m/ \hbar^2 $  which determines $\tau_{sc}=2 \hbar/ \Gamma_0^0$ in
Eq.~(\ref{tautrans}), and the impurity range $r_0$. In the presence of a
strong perpendicular magnetic field, these parameters enter the equations for
the Green's functions and the conductivities via the energy
$\Gamma_0=\sqrt{\Gamma_0^0 \, \hbar \omega_c} =\sqrt{n_i}\, u_0/\ell\equiv
\gamma \sqrt{B_0}$, which determines the scattering-induced broadening of the
spectrum, and the length ratio $r_0 /\ell$, which determines 
the anisotropy of the impurity scattering. \cite{Ando82:437} This model allows
a direct comparison 
between the results for point-like ($r_0=0$) and finite-range ($r_0\neq
0$) impurities.  Major simplifications of the transport calculation
occur only for the unmodulated system and $r_0=0$.  A direct inspection
of Eqs.~(\ref{greenf})-(\ref{vertexc}) shows that in this case the
self-energy operator becomes a simple (energy-dependent) c-number,
and the current vertex corrections, Eq.~(\ref{vertexc}),  
vanish. \cite{Ando82:437,Gerhardts71:126}

\subsection{The usual `c-number approximation'}
Although this is no longer true in the presence of a modulation, even
if $\delta$-impurities are assumed, the ansatz that the self-energy is still
a c-number (i.e., independent of all the quantum numbers), has been
used quite often.\cite{Zhang90:12850,Tornow96:16397,Steffens98:3859,Tan94:1827}
A justification of the c-number approximation (CNA) may be  that,
besides its simplicity, it is exact (for $\delta$-impurities) in the limit of
zero modulation and, therefore, may be 
reasonable also for sufficiently weak modulations. This ansatz is sufficient
to ensure the vanishing of the vertex corrections, but, on the other hand,
it captures  essential effects
of the collisions and of the density of states.\cite{Zhang90:12850}
Moreover, it is regarded as satisfactory if one is interested mainly in the
influence of the system geometry and not of impurity scattering on the
transport properties.
In the CNA the Green function operator is diagonal in the modulated basis, 
$(nX_0|G^{\sigma}(E)|n'X_0) =\delta_{n,n'}  G^{\sigma}_{nn}(X_0,E)$,
such that
the longitudinal conductivities may be written as
\begin{equation} 
\sigma_{\alpha\alpha}(E)=\frac{\hbar e^2}{\pi^2\ell^2}
\int_0^a \frac{dX_0}{a} \sum_{nn'} |(nX_0|v_{\alpha}|n'X_0)|^2 \
{\mathrm Im} G^-_{nn}(X_0,E) \ {\mathrm Im} G^-_{n'n'}(X_0,E) \,. 
\label{condcna} 
\end{equation} 
In this expression the essential difference between $\sigma_{xx}$
and $\sigma_{yy}$, generated only by the anisotropy of the modulation, is
determined by the diagonal ($n=n'$) contribution which exists only
for $\alpha=y$, since $(nX_0|v_x|nX_0)=0$.  This intra-Landau-band
contribution, also known as {\em band conductivity}, \cite{Azin84:1469}
is related to the net motion of electrons in the direction perpendicular
to the electric field and/or the gradient of the magnetic field defining the
modulation, with the group velocity
\begin{equation}
(nX_0|v_y|nX_0)=-\frac{1}{m\omega_c}\frac{d E_{n,X_0}}{dX_0}\,.
\label{driftve} 
\end{equation} 
If the scattering broadening $\Gamma_0$ is much smaller than the modulation-induced
bandwidth of the Landau levels, and if this in turn is so small that adjacent
Landau bands do not overlap, simple estimates of the dependence of the
different conductivity contributions on $\Gamma_0$ and
the DOS at the Fermi level, $D(E_F)$, are available. As shown in  Appendix~B,  
the band conductivity diverges for $\Gamma_0^0 \rightarrow 0$
 like $[\Gamma_0 D(E_F)]^{-2}$ (provided $E_F$ is inside a Landau band and not
 too  close to a band edge).  This explains -- in the quantum
treatment -- the Weiss oscillations, with minima when $E_F$ intersects
a flat band.\cite{Zhang90:12850}  The inter-Landau-band contribution
to Eq. (\ref{condcna}) ($n\neq n'$), also known as {\em scattering
conductivity}, in general behaves like $[\Gamma_0 D(E_F)]^{2}$,
and entirely determines $\sigma_{xx}$.  Within the CNA, for a sufficiently
clean system (small $\Gamma_0^0$), the scattering conductivity 
$\sigma_{yy}^{scat} $
is much smaller than the band conductivity at weak magnetic fields, i.e.,
in the regime of the Weiss oscillations, and both conductivity contributions
may become comparable 
for strong fields, in the regime of the Shubnikov - de Haas (SdH)
oscillations. Also, the scattering conductivities are almost the same
in the modulated and unmodulated directions.\cite{Manolescu97:9707}

For large $B_0$, the estimates given in Appendix~B yield $\sigma_{yy}^{band}
\sim (B_0)^{-2}$ for pure electric and $\sigma_{yy}^{band}
\sim (B_0)^{-4}$ for pure magnetic modulation, leading to  $\Delta
\rho _{xx}/ \rho_0 \sim (B_0)^2$  and  $\Delta
\rho _{xx}/ \rho_0 \sim (B_0)^0$, respectively. All these results for the
CNA are in agreement with the corresponding classical results summarized in
Sec.~\ref{classresults}. This confirms our expectation, that consideration of
vertex corrections is important, 
not only from a quantitative point of view, in order to
include the effect of anisotropic scattering, 
but also from a qualitative point of view, to
avoid violation of the equation of continuity.
The neglect of vertex correction, e.g., in the CNA,  leads to uncontrolled and
(at least for large $B_0$) unacceptable
results,  even for isotropic scattering.\cite{Menne98:1707} 
The deficiencies of the CNA also became clear from a quantitative comparison
between experimental results and quantum-mechanical calculations which were
non-perturbative with respect to the periodic modulation
potential.\cite{Steffens98:3859} In that work it turned out to be impossible to
fit the magnitude of the experimental resistance and the dominance of
scattering conductivity over band conductivity (indicated by resistance {\em 
minima} at the flat band conditions) at the same time.

\subsection{Beyond the CNA}
The aim of the present paper is to go beyond the CNA. Indeed, there exists no
justification of the CNA within the formalism of the SCBA. Even if we {\em
  assume} that the Green operator (\ref{greenf}) is diagonal, say in the
modulated basis, 
and insert that into Eq.~(\ref{scba}), the evaluation of the kernel (see
Appendix A) yields a non-diagonal self-energy, and thus in the next iteration
step a non-diagonal Green operator.  One finds, that there exists no basis in
which the Hamiltonian of the modulated system (without disorder), the
impurity-averaged Green function, and the self-energy operator are
simultaneously diagonal, since these three operators do all not commute with
each other. 
For obtaining numerical solutions of the self-consistent equations
(\ref{greenf})-(\ref{scba}) and (\ref{vertexf})-(\ref{vertexc}) we find it
convenient to express the matrix elements of all the operators in the
Landau basis, since the kernels describing the impurity-averaging in the
SCBA are independent of the modulation in this case.

Then, for a given modulation model and for fixed values of  energy $E$ and
average magnetic field $B_0$,
 the matrix elements of the Green operator $\langle m,X_0 |G^\sigma
(E)|n,X'_0 \rangle = \delta_{X'_0,X_0} G^\sigma_{m,n}(X_0,E)$ depend on two
discrete (Landau) quantum numbers and a quasi-continuous one (center
coordinate). Equation~(\ref{scba}) provides a linear relation between these
and the self-energy matrix elements, which have the same structure, mediated
by the SCBA kernel, which depends on four 
discrete indices and  a single continuous one, since the $X_0$ relation is of
a convolution type (see Appendix A). Thus, it is convenient to expand the
$X_0$ dependence into a Fourier series. Moreover, we restrict our consideration
to modulation models of defined parity, which allows to reduce the number of
Fourier coefficients by a factor of two. Eqs.~(\ref{greenf})-(\ref{scba})
provide a non-linear integral equation for $G^\sigma_{m,n}(X_0,E)$ or,
equivalently, $\Sigma^\sigma_{m,n}(X_0,E)$. At finite temperature, the
solutions are needed in a finite interval around the chemical potential, which
has to be calculated for given average electron density. Having calculated
these solutions, we can solve the linear integral equations
(\ref{vertexf})-(\ref{vertexc}) for the vertex functions, which have a similar
matrix structure as the Green operator and are needed to calculate the
conductivity components. It is evident that the requirements on storage
capacity and 
computational time increase rapidly with the number of Landau levels and of
Fourier coefficients, which have to be taken into account.  Due to the limited
computer facilities, we will therefore have to use somewhat unrealistic model
assumptions, which allow to restrict, e.g., the number of necessary Fourier
coefficients. We will also restrict the  investigation of scattering-anisotropy
effects to the consideration of only two values of the impurity range $r_0$.
  The results of the
CNA can be found by taking $r_0=0$ and,  simultaneously, reducing all the
Fourier series to the first term with $p=0$ (which is the average value, see
Appendix A). Note, that the latter prescription is an additional approximation
which has no justification, but allows the
comparison of the correct results with those of the CNA.

We shall consider the effects of the anisotropic collisions on the scattering
and band contributions to the diagonal components of the {\em conductivity
  tensor}. In order to illustrate the results for
the {\em resistivity tensor}, Eq.~(\ref{rhotensor}), we shall simplify the Hall
conductivity by completely neglecting the impurity effects on it.  Of course,
quantum Hall plateaus are thus ignored, although they are seen in the
experiments on modulated systems for larger magnetic fields. However, we assume
that the structure
of the longitudinal resistivities is mostly determined by the periodic
Landau bands for sufficiently low magnetic fields, and not by localization
effects which are beyond the SCBA. 
Especially the anisotropy effects corresponding to the band conductivity
are expected to result from the diagonal components of the conductivity
tensor (a corresponding diagonal contribution in the modulated
basis does not exist for the Hall conductivity).
With these assumptions, 
the Hall conductivity is given by
\begin{equation}
\sigma_{xy}= \frac{2 i\hbar e^2}{\pi\ell^2}
\int_0^a \frac{dX_0}{a} \sum_{n\neq n'} {\cal F}(E_{nX_0})
\,\frac{(nX_0|v_{x}|n'X_0) (n'X_0|v_y|nX_0)}
{(E_{nX_0}-E_{n'X_0})^2} \,. 
\end{equation}

\section{Results}

For computational reasons we have chosen our modulation parameters
such that we can obtain convergent results by truncating the Landau
basis to the first 10-30 wave functions, and the Fourier series with
respect to $X_0$ to the first 10-20 terms.  We avoid the necessity
of including higher Fourier coefficients by keeping a relatively weak
modulation amplitude, and also by choosing $r_0$ shorter, but comparable
to $\ell$.  For a disorder broadening much smaller than the typical
bandwidth, the periodic matrix elements $G^{\sigma}_{nn'}(X_0,E)$ and
$F^{\sigma\sigma'}_{nn'}(X_0,E)$ as functions of $X_0$ exhibit large pole-like
structures (which become poles for $\Gamma_0\to 0$),
 if $E$ falls into the spectrum $E_{m,X_0}$ of the Landau bands
$m=n$ or $m=n'$.  Apparently
this structure requires the inclusion of many Fourier coefficients.  But,
due to the Gaussian factors of the Landau wave functions, the matrix
elements of the collision operator $\Gamma^2$ decay like Gaussians with
increasing $|X_0-X_0'|$, Eq.~(\ref{gammam}), and thus the higher-order
Fourier amplitudes have a vanishing (also Gaussian) contribution to
$\Sigma^{\sigma}_{nn'}(X_0,E)$ and $I[F^{\sigma\sigma'}_{nn'}(X_0,E)]$,
respectively.  This can be directly seen on the analytical results
available for $r_0=0$, Eqs.~(\ref{integd0}) and (\ref{efn}).  For $r_0\neq
0$ the Fourier coefficients are calculated by integrating the periodic
functions, and in order to keep a reasonable number of points in the
integrals we restricted ourselves - nevertheless - to a relatively large
disorder broadening, i.e., $\Gamma_0$ smaller, but comparable to the width
of the Landau bands.  Therefore, we do not attempt to give necessarily
realistic results, but rather to identify and to understand the effects
of the anisotropic scattering qualitatively.

%

\subsection{Electric modulation}

We start with an example of a pure electric modulation ($B(x)\equiv
B_0$), determined by the electrostatic potential $V(x)=U\cos(2\pi x/a)$,
with $U=0.8$ meV and $a=400$ nm. The material parameters are for GaAs:
$m=0.067m_0$, the electron density ($n_{\mathrm el} =1.93 \cdot
10^{11}\,$cm$^{-2}$, $E_F(B_0$=$0)=6.91\,$meV) 
 chosen such that $\nu B_0=8$ T, $\nu$
being the filling factor, and we assume spin degeneracy.  We fix the
temperature to 1 K, and the disorder parameter $\gamma=0.5$ meV T$^{-1/2}$
corresponding to a mobility $\mu = 2.2 \cdot 10^5$ cm$^2/Vs$ at $B_0=0$ in CNA
and to an increasingly larger mobility with growing anisotropy of scattering as
the ratio $\tau_{tr} / \tau_{sc}$ increases.

In Fig.~2 we show $D(E_F)$ and a typical energy spectrum, with $B_0=0.81\,$T
so that the Fermi level is in the Landau bandwidth $n=4$.  
We compare the CNA with the results of the calculations with a matrix
self-energy for $r_0=0$ (delta impurities) and for a finite $r_0$.
Broadened van Hove singularities (VHS's) are  resolved in
all cases.  The small maximum at $B_0=0.9\,$T is an artifact due to the
elliptic shape of the DOS, typical for the SCBA: unlike in the CNA, in the
other calculations the self-energy depends on the center coordinate,
and since the disorder broadening is comparable to the bandwidth, the  two
high-DOS peaks due to the band edges may partially overlap, yielding extra
maxima in between. Such 
details are, however, not important for the conductivities.

Results for the conductivities are shown in Fig.~3, again in the CNA,
and also with a matrix self-energy (including the vertex corrections), 
for two choices of $r_0$.
Note, that the Landau bands do not overlap in the present case. Thus, in the
CNA the scattering and band conductivities are given by the estimates
Eqs.~(\ref{siscatt}) and (\ref{siband2}) in Appendix B. Numerically,
we obtain $\sigma_{xx}\approx \sigma_{yy}$ in the CNA,
which tells that, {\em in this approximation} and for the chosen values of
modulation amplitude and (relatively large) disorder broadening, 
 the band conductivity is very small.
The first flat-band condition (see Appendix B)
%
%
at $E_F$ is obtained for $B_0\approx
0.5\,$T, where the scattering conductivities have maxima.  When the vertex
corrections are included, even for $r_0=0$, the situation completely
changes: with increasing magnetic field the two conductivities have
opposite evolutions.  The pure scattering conductivity $\sigma_{xx}$
is drastically reduced, and the band conductivity leads to  a dramatic
increase of $\sigma_{yy}$.  While the double-peak structure of the DOS is no
longer resolved in $\sigma_{xx}$, it may survive in $\sigma_{yy}$.

We can interpret these results in physical terms, taking into account that the
anisotropy of the modulation results in anisotropic electron states: unbound
propagation in the $y$ direction and confined cyclotronic motion in the $x$
direction within the cyclotron diameter $2R_c=2\ell\sqrt{2n+1}$.  For low
magnetic fields such that $2R_c\gg a$ the effect of the modulation on the
scattering events is averaged out and the scattering remains isotropic for
$r_0=0$. This changes for $2R_c\ll a$, when the electrons propagate along
relatively narrow channels in the $y$ direction.  The {\em decrease} of
$\sigma_{xx}$, with respect to the CNA, can be understood just like the effect
of the scattering anisotropy on the longitudinal (scattering) conductivities 
in the absence of the modulation,  but in the presence of the magnetic field.
\cite{Ando82:437} At the same time, the {\em increase} of the band conductivity
contributing to $\sigma_{yy}$  can be understood in analogy to the effect of
the scattering anisotropy on the conductivity of quasi-free electrons, with no
modulation and no magnetic field.
A Landau band has the strongest dispersion in its center and an uncertainty in
energy due to disorder broadening translates into a much smaller range $\Delta
k_y \sim \Delta X_0$ of available final states after scattering as compared to
the band edges. This in turn leads to a stronger increase in forward scattering
and, thus, to the enhanced band conductivity in the middle of a band. Obviously,
this effect is stronger for finite-range impurities than for $r_0=0$.
The decomposition of $\sigma_{yy}$ into scattering and band contributions
is however complicated by the vertex corrections, which in fact with
increasing modulation amplitude mix them increasingly together.  Therefore the
behavior of the scattering contribution to $\sigma_{yy}$ cannot be very 
clearly identified and understood with these simple interpretations. 

We found numerically that,
for the results shown in Figs.~3 and 4, the vertex corrections corresponding
to $I[F_{\alpha}^{++}]$ and $I[F_{\alpha}^{--}]$ are negligible.  For delta
impurities one can show analytically that they vanish exactly for
$\alpha=x$ (see Appendix A).  In general, their small values can
be explained with the help of the Ward identities, which relate
$I[F_{\alpha}^{\sigma\sigma}]$ to the commutators of the self-energy
with the position operators.\cite{Gerhardts71:126} The matrix elements
of the vertex corrections become determined by differences between
off-diagonal matrix elements of the self-energy, which are small
unless the Landau-level mixing is very strong.  So practically all the
anisotropy effects obtained for the electric modulation are determined
by $I[F_{\alpha}^{-+}]$.

In general, the resistivities reproduce the structure of the
conductivities, according to Eq.~(\ref{rhotensor}) where the denominator
is in fact given by the square of the Hall conductivity.  $\rho_{xx}$
is qualitatively similar to $\sigma_{yy}$, and $\rho_{yy}$ to
$\sigma_{xx}$ respectively. Qualitatively, the results confirm the
expectations, which we deduced from the analysis of the classical and the CNA
results:  
anisotropic scattering reduces the transport scattering rate and thus reduces
the scattering conductivity  (and consequently $\sigma_{xx}$)
and increases the band conductivity (and thus  $\sigma_{yy}$).  For
$\Gamma_0^0 \ll \hbar \omega_c$, this should enhance $\rho_{xx}$,
see Fig.\ 4 and Fig.\ 5(a), but lower $\rho_{yy}$, see Fig.\ 5(b). 
In particular, for $r_0=0$, the strong enhancement of $\rho_{xx}$
over the CNA result observed at larger $B_0$ values in Fig.\ 4, also
agrees with the predictions based on Boltzmann's equation: the neglect of
backscattering (i.e., vertex corrections, as in the CNA), which violates
the equation of continuity, results in too small values for  $\rho_{xx}$
and becomes increasingly important  with increasing $B_0$.

\subsection{Magnetic modulation}

In the following examples we consider a pure magnetic modulation
($V(x)\equiv 0$), defined by a periodic magnetic field of the form
$B(x)=B_0+B_1\cos(2\pi x/a)$.  We shall discuss two cases, first with a
high and then with a low band conductivity, in order to separate the
anisotropy effects, and also to compare with the available experimental
results.

\subsubsection{The regime of high band conductivity}

We choose the modulation parameters comparable to those in the experiment
by Ye et.\ al.\ on strongly magnetically modulated 2DEG. \cite{Ye98:330}
The experiment shows huge oscillations of the magnetoresistivity
$\rho_{xx}$ for values of the average magnetic field $B_0$ in the
intermediate regime between that of positive magnetoresistance and
that of distinct SdH peaks.  The positive magnetoresistance effect
is also huge and covers the Weiss oscillations.  In general,
for both electric and magnetic modulations, the positive
magnetoresistance and the Weiss oscillations have been successfully
explained by the classical transport calculations, where the
classical analog of the band conductivity is in fact calculated.
\cite{Beenakker89:2020,Gerhardts96:11064,Menne98:1707,Mirlin98:12986}
In the intermediate regime mentioned here, the quantum effects become
important. The modulated magnetic field is strong and generates
Landau bands that partially overlap, yielding oscillations of the DOS
at the Fermi level.  We ascribe the huge oscillations observed in the
experiment to the resulting oscillations of the band conductivity.

In our calculations we take $a=400$ nm and $B_1=0.23$ T for the modulation,
and $\gamma=0.3$ meV T$^{-1/2}$ 
for the scattering-energy parameter.  In Fig.~5 we
show the calculated resistivities with $B_0$ starting at the end of
the positive-magnetoresistance regime.  The vertex corrections are
included, both for zero and finite impurity range.  The contribution of
the scattering conductivity is negligible in $\sigma_{yy}$, and thus
in $\rho_{xx}$ [see Eq.\ (\ref{rhotensor})],
because the Landau bands are much wider than the
disorder broadening $\Gamma_0$.  The filling factor varies between 21
and 11.4, and the number of Landau bands intersected by the Fermi level,
for a  fixed $B_0$, varies between four and two. 

In Fig.~6(a)  we have selected three energy spectra with a VHS at the
Fermi level, corresponding to a Landau band edge.  The singularities are
well resolved in $D(E_F)$, Fig.~6(b). The profile of $D(E_F)$ is very well
reproduced in both resistivities calculated at zero temperature, Fig.~5.
$\rho_{xx}$ has minima and $\rho_{yy}$ has maxima where $D(E_F)$
has maxima.  This clearly tells that $\rho_{xx}$ is determined by the
band conductivity and $\rho_{yy}$ by the scattering conductivity.  For a
finite temperature, like 1 K, the VHS are smeared and the resistivities
oscillate according to the number of the Landau bands at the Fermi level.
For instance, for 0.442 T $< B_0 <$ 0.473 T {\em three} Landau bands are
intersected by the Fermi level, while for 0.473 T $< B_0 <$ 0.504 T {\em
four} bands, Fig.~6(a), such that $\rho_{xx}$ has a maximum in the first
interval (low $D(E_F)$) and a minimum in the second one (high $D(E_F)$),
while $\rho_{yy}$ has the opposite evolution.

Just like in the case of the electric modulation, the effect of the
scattering anisotropy and of the vertex corrections is to increase the
band conductivity and to decrease the scattering contributions.  Here,
such effects are not so important for $r_0=0$, due to the relatively low
average magnetic field. Therefore the DOS-induced oscillations of the
resistivities can be qualitatively explained even in the CNA (not shown in
Fig.~5; compare the estimate for the band conductivity in the case of
overlapping bands, Eq.~(\ref{overlap}) in Appendix B), and the oscillations
can be amplified by reducing the disorder
parameter $\gamma$.  Of course, quantitative agreement with the experiment
cannot be expected in our calculations, where the disorder parameters
are chosen according to computational possibilities rather than to
experimental requirements.  The model of
Gaussian impurities involves two parameters, $r_0$ and $\gamma\sim
\sqrt{n_i}u_0$. 
For a direct comparison between the theoretical and the experimental
results, for different choices of $r_0$ one should also adjust $\gamma$,
e.g.\ by imposing a fixed zero-magnetic-field resistance. This is given by the 
transport time $\tau_{tr}$, Eq.~(\ref{tautrans}), which can be calculated
analytically for our Gaussian impurity model.
To keep $\tau_{tr}$ fixed, we should compare 
the results for $r_0=10$ nm and $\gamma=0.3$ meV T$^{-1/2}$ 
with those for $r_0=0$ and $\gamma\approx 0.05$ meV T$^{-1/2}$. Of 
course, the realistic scatterers have a long range, with Coulomb tails
determined by the spacer width, and may be not too well approximated by
Gaussians. 

Experimental results in this regime exist only on the resistivity component
$\rho_{xx}\, $, not on $\rho_{yy}$, to our knowledge. 
We  predict that, as indicated by our
calculations, $\rho_{yy}$ should be at least one order of magnitude
smaller than $\rho_{xx}$ and should oscillate in a similar manner, but
with minima and maxima interchanged.  Unlike the Weiss oscillations, which are
determined by the geometric commensurability of the cyclotron diameter with
the modulation period, \cite{Weiss89:179} these novel modulation-induced
 oscillations occur
for Fermi energies below the lowest commensurability (flat-band) condition,
and are essentially a DOS effect.

\subsubsection{The regime of low band conductivity}

Measurements of the magnetoresistance in both $x$ and $y$ directions,
in the quantum regime, have been recently performed on magnetically modulated
2DEGs in InAs heterostructures. \cite{Hthesis,Heisenberg} The electron
density is typically one order of magnitude higher than in GaAs systems,
and thus the filling factors are much larger.  The SdH peaks observed
in $\rho_{xx}$ are about 20 times higher than in $\rho_{yy}$.  At the
same time, with decreasing the uniform field $B_0$, the SdH minima
at even filling factors are replaced by resistivity maxima, while
the minima shift to odd filling factors. It has been convincingly
demonstrated, that these minima are not due to spin splitting, which in fact
is not resolved in this case.  A similar even-odd shift has been observed in
electrically modulated systems, and explained by the behavior
of the scattering conductivity when two Landau bands partially overlap.
The overlapping  VHS of adjacent Landau bands yield maxima instead of SdH
minima, while minima occur 
for $E_F$ in the middle of a band (lower DOS).  \cite{Tornow96:16397} We
therefore believe that, similarly, also in the experiment of Heisenberg et.\
al., \cite{Hthesis,Heisenberg} only the scattering conductivities are observed, 
and thus the big difference between $\rho_{xx}$ and $\rho_{yy}$ is not
due to the band conductivity, but due to an anisotropy of the scattering
contributions to  $\rho_{xx}$ and $\rho_{yy}$.

We use the parameters of InAs, $m=0.023\, m_0$, an electron density such
that $\nu B_0=75\,$T, and spin degeneracy. For the magnetic modulation
we take $a=300$ nm and $B_1=0.1\,$T.  We calculate the resistivities in the
regime of the SdH oscillations for impurity parameters suitable to obtain
dominant contributions from the scattering conductivities: $r_0=5$ nm
and two values for the disorder broadening, $\gamma=5$ 
meV T$^{-1/2}$ and $\gamma=6$ meV T$^{-1/2}$ .  
The resistivities are shown in Fig.~7. At $T=4.2\,$K, the thermal energy
$k_BT$ is much smaller than the width of the Landau bands. A reference
energy spectrum is displayed in Fig.~8.  In the explored interval
of $B_0$ values, the cyclotron diameter of the states at the Fermi level varies
between $a/2$ and $a/3$, such that the vertex corrections are very strong.

The contribution of the band conductivity is seen as a central maximum in each
of the first three SdH peaks of $\rho_{xx}$, located in $2.5\,$T$ < B_0
< 3.15\,$T. With increasing $B_0$ these central structures disappear,
apparently because the effective disorder broadening $\Gamma_0$ increases.
Indeed, by increasing the parameter $\gamma$ the mentioned central
maxima decrease, and at the same time the lateral shoulders of $\rho_{xx}$,
as well as $\rho_{yy}$, increase.

This behavior clearly shows that the scattering conductivity is
dominant here. The anisotropy of the scattering conductivities, i.\
e.\ $\sigma_{xx}\ll \sigma_{yy}$, or $\rho_{xx}\gg \rho_{yy}$, is
an effect {\em qualitatively} determined by the vertex corrections,
and not by the band conductivity.  We have already mentioned such an
effect in the discussion of Figs.~3 and 4, where it was less pronounced
and combined with the presence of a higher band conductivity,
but visible even for delta impurities.  This type of anisotropy,
of the scattering conductivities, cannot be obtained in the CNA.
In that approximation again $\sigma_{xx}\approx \sigma_{yy}$ (or
$\rho_{xx}\approx\rho_{yy}$), and unlike in the example of the electric
modulation, the CNA resistivities  become here several times bigger
than those plotted in Fig.~7.  Our results for the magnetic modulation
thus show characteristic similarities with  the experimental results
of Heisenberg et.\ al. We are not able to offer a simple argument that
could explain the strong anisotropy of the scattering conductivities
obtained from the numerical calculations quantitatively. However, we
attribute this anisotropy  to both the anisotropic electronic states and
the anisotropic impurity scattering.  Of course, strong anisotropies can
be understood qualitatively in the physical picture of quasi-classical
orbits with a relaxation of the rigid energy-versus-$X_0$ dispersion
due to disorder broadening (see the discussion in section IV.A of
the enhancement of the band conductivity in the middle of a Landau
band). Scattering processes that lead to an effective motion in $y$
direction, and thus contribute to $\sigma_{yy}^{scat}$, are possible
with arbitrarily small changes of the center coordinate $X_0$, and are
favorized if forward scattering predominates. On the other hand, elastic
scattering processes leading to a substantial effective motion in $x$
direction require a finite (large) change of the $X_0$ coordinate, i.\ e.\
a scattering by a large angle. Such processes, and consequently their
contributions to $\sigma_{xx}^{scat}$, are suppressed by predominant
forward scattering, and for $2R_c<a$ they become even more improbable,
since they require a tunneling through an energy barrier.  These aspects
have to be better clarified by future analytical calculations.

An analogous set of experimental results has been obtained for an electric 
modulation of a very short period (32 nm). \cite{Petit}  The resistivity
tensor is anisotropic, the resistance in the modulated direction being up
to five times bigger than in the uniform direction.  The amplitudes of the
Shubnikov - de Haas peaks are well represented by Dingle plots, which are
specific to 
unmodulated (homogeneous) systems, where only the scattering conductivity
exists.  The results  suggest that the band conductivity plays no role
in these data, and thus the resistance anisotropy is determined
by the scattering anisotropy, as we find in our present calculations.

\section{Conclusions and final remarks}

We have shown with numerical calculations that for strong magnetic
fields the anisotropic character of the electron-impurity scattering has
both quantitative and qualitative effects  on the resistivity tensor of
modulated systems.
Our magnetotransport calculation is based on a consistent evaluation of the
disorder broadening of the single-particle Green-functions (and thereby the
density of states) and of the Kubo formulas for
the conductivity tensor. Formally, the anisotropy effects enter through vertex
corrections, which are calculated from the Bethe-Salpeter (integral) equations.
In contrast to homogeneous systems, for modulated systems these anisotropy
effects are important even for short-range 
($\delta$-potential) impurities, if the average magnetic field is so strong
that the cyclotron diameter of the electrons
at the Fermi level is smaller than the modulation period.  Of course,
such effects become even much more important for impurities of finite range. 

The main results of including the anisotropy effects (vertex corrections)
are:\\ 
%
(1) {The band conductivity increases, just like the conductivity of a
    homogeneous 2DEG for zero magnetic field (which is inversely proportional
    to the disorder-scattering rate).} \\
(2) {The scattering conductivities decrease, just like the longitudinal 
    conductivities of a homogeneous 2DEG in a high magnetic field (which are
    proportional to the  disorder-scattering rate).}\\
(3) {The scattering conductivity in the direction of the modulation 
    becomes lower than that in the direction perpendicular to the modulation.}
%

The comparison with the available experimental results suggests that the
scattering anisotropy is important not only in the classical regime 
(low $B_0$) \cite{Menne98:1707,Mirlin98:12986}, but also in the quantum
case (high $B_0$). This is seen, in both regimes, in the band conductivity, 
which has a classical origin, but also in the pure quantum scattering 
conductivity. Experimental results for both $\rho_{xx}$ and
$\rho_{yy}$ in a unidirectional modulation, for high magnetic fields, 
are however rarely available,\cite{Petit,Hthesis,Heisenberg} but still, 
our results agree qualitatively with these experiments.

The numerical calculations of this work require the solution of the
non-linear Dyson equation for the Green operator, and of the linear
Bethe-Salpeter equation for the vertex operators, each of which depends
on two discrete (Landau) quantum numbers and one quasi-continuous ($X_0$)
quantum number, and in addition on the continuous parameters energy $E$
and $B_0$. To make this complex problem tractable with our numerical
capabilities, we had to make rather restrictive assumptions for the
model parameters.  In particular, we assumed a relatively large impurity
broadening to enssure a rapid convergence of the Fourier expansions.
Therefore, the results shown in this paper are based on a few examples,
rather than on a complete insight into the structure of the vertex
corrections and a systematic analysis of the length scales imposed by
the modulation period, cyclotron radius, and impurity range.

A more detailed, systematic investigation of anisotropic-scattering effects
in more realistic cases of high-mobility modulated systems exceeds
our present possibilities concerning computer memory and CPU time.
Although quantitatively our results depend on the parameters of
the Gaussian model for impurities, we believe they do not change
qualitatively when the impurity range is increased to more realistic
values to simulate the situation at high-mobility semiconductor
interfaces.  Of course, in that case the realistic impurity model is
the  Coulomb potential of a charged particle situated outside the 2DEG.
Nevertheless, for such a model, in order to avoid singularities in
the scattering integrals (\ref{gammam}) or (\ref{integd}), one has to
include the screening effects of the 2DEG on the impurity potential,
e.g.\  within a Thomas-Fermi approximation. Note, that the screening of
the periodic lateral potential in the case of the electric modulation
may considerably reduce the width of the Landau bands, and thus the
influence of VHS on the magnetoresistivities, which we do not expect
for a magnetic modulation.  \cite{Manolescu97:9707,Gossmann98:1680} On
the other hand, the dependence of the screening length on the DOS at the
Fermi level, and also the alternating of quasi-metallic (compressible)
and insulating (incompressible) phases in strong magnetic fields, open
the possibility of describing more complicated, but qualitatively new
effects in the resistivities.

\acknowledgements

A.M. has been supported by Universit\"at Regensburg, under
Graduiertenkolleg "Komplexit\"at in Festk\"orpern" (GK176).  
Useful discussions with David Heisenberg, Dieter Weiss, and
Bernard Etienne are also acknowledged.

\appendix

\section{The SCBA kernel} \label{scbakern}

In the numerical calculation we find it most convenient to use the Landau 
basis,
\begin{equation}
\langle x,y
|\, nX_0\rangle=\exp(-iX_0y/\ell^2)\phi_n[(x-X_0)/\ell]/(L_y\ell)^{1/2} \,,
\label{landaub}
\end{equation}
with $n=0,1,2,...$ the orbital (oscillator) quantum number and 
$X_0=2\pi\ell^2 \times (integer)/L_y$ the (quasi-continuous) center 
coordinate.
Eqs. (\ref{scba}) and (\ref{vertexc}) have the same structure, 
which in matrix form reads
\begin{equation}
A_{nm}(X_0)=\sum_{n'm'X_0'} \Gamma_{nm,n'm'}^2 (X_0,X_0') B_{n'm'}(X_0')\,,
\label{formac}
\end{equation}
$A$ being either $\Sigma$ or $I[F]$, and $B$ being either $G$ or $F$. Due to
the translational invariance along $y$,  all these operators are diagonal in
$X_0$ and, similar to the energy spectrum $E_{n, X_0}$, the  $X_0$-dependence
is periodic with the modulation period $a$.  In the Landau basis the matrix
elements of the collision 
operator depend only on the difference of the center-coordinates,
\begin{equation}
\Gamma_{nm,n'm'}^2 (X_0-X_0')= n_i\int d{\bf R}\
\langle n X_0| u({\bf r}-{\bf R})|n'X_0'\rangle\
\langle m' X_0'| u({\bf r}-{\bf R})|m X_0\rangle \,,
\label{gammam}
\end{equation}
as can be shown, e.g., by using the Fourier transform $\tilde u$ of $u$.
Hence, Eq. (\ref{formac}) becomes a convolution with 
respect to $X_0'$, and it can be simplified by the Fourier-series
expansions
\begin{equation}
B_{n'm'}(X_0')=\sum_{p\geq 0} b_{n'm'}(p)\
\cos \left[pKX_0'+(n'-m'+j)\frac{\pi}{2}\right]\,,
\label{coeff}
\end{equation}
where $K=2\pi/a$ is the modulation wave vector, and
where the phase shifts are chosen to satisfy the reflection symmetry 
imposed by the modulations: we take $j=0$ for $B=G$, because
$G_{nm}^{\sigma}(-X_0)=(-1)^{(n-m)}G_{nm}^{\sigma}(X_0)$, and 
$j=1$ for $B=F$, because
$F_{nm}^{\sigma\sigma'}(-X_0)=(-1)^{(n-m+1)}F_{nm}^{\sigma\sigma'}(X_0)$.
These properties follow for our simple cosine-modulations from the
corresponding properties of the matrix elements of the Hamiltonian and
 of the velocity operator, respectively. Then,
Eq. (\ref{formac}) becomes
\begin{equation}
A_{nm}(X_0)=\sum_{n'm'}\sum_{p\geq 0} b_{n'm'}(p) D_{nm,n'm'} (pK\ell)
\cos \left[pKX_0+(n-m+j)\frac{\pi}{2}\right]\,,
\label{formacf}
\end{equation}
where
\begin{equation}
D_{nm,n'm'} (pK\ell)=\frac{n_i}{\pi \ell^2}
\int_0^{\infty} dq\, q \ [\tilde u(\frac{q\sqrt{2}}{\ell})]^2 \,
E_{n n'}(q^2) \, E_{m m'}(q^2) \,
J_{n-n'-m+m'}(pK\ell q\sqrt{2}) \,,
\label{integd}
\end{equation}
with
\begin{equation} \label{enmvonz}
E_{nm}(z)=\left(\frac{m!}{n!}\right)^{1/2} z^{(n-m)/2}
e^{-z/2} L_m^{n-m}(z)= (-1)^{m-n}E_{mn}(z)\,,
\label{efn}
\end{equation}
$L_n^m(x)$ being the Laguerre polynomials, and $J_n(x)$ the Bessel
functions.  We have the following properties:
\begin{equation}
D_{nm,n'm'}=(-1)^{n+m+n'+m'} D_{mn,m'n'} \,,
\end{equation}
\begin{equation}
D_{nm,n'm'}= D_{n'm',nm} \,.
\end{equation}

In particular, for delta impurities, i.e., for $\tilde u (q) \equiv u_0$,
Eq.~(\ref{integd}) can be integrated analytically, and one obtains
\cite{Gradshteyn}
\begin{equation}
D_{nm,n'm'}(pK\ell)=\frac{n_iu_0^2}{2\pi\ell^2} \, E_{nm}(p^2z) \,
E_{n'm'}(p^2z)\,, 
\label{integd0}
\end{equation}
where $z=(K\ell)^2/2$.

Using the Fourier series (\ref{formacf}) in Eqs.~(\ref{scba}) and
(\ref{vertexc}), we solve iteratively the self-consistent equations
(\ref{greenf})-(\ref{scba}) and (\ref{vertexf})-(\ref{vertexc}), starting
from the CNA.  The velocity matrix elements, used in Eqs.~(\ref{vertexf})
and (\ref{kubof}), are
\begin{equation} \label{velomat}
\langle n' X_0 |v_{\alpha} | n X_0\rangle =
\eta_{\alpha}\frac{\ell\omega_c}{\sqrt{2}}
\left(\sqrt{n+1}\ \delta_{n',n+1}+
\eta_{\alpha}^2\sqrt{n'+1}\ \delta_{n,n'+1}\right)\,,
\end{equation}
with the notation $(\eta_x,\eta_y)=(i,1)$.

One can directly show the following symmetries:
\begin{equation}
(F_{\alpha}^{\sigma\sigma'})^*_{mn}=(F_{\alpha}^{-\sigma'-\sigma})_{nm}\,,
\end{equation}
for both $\alpha=x,y$, and from the symmetries of the Hamiltonian, which in
the Landau representation is a real, symmetric matrix, and of the velocity
operators, see Eq.~(\ref{velomat}),  
also
\begin{equation}
(F_{x}^{\sigma\sigma'})_{mn}=-(F_{x}^{\sigma'\sigma})_{nm}\,,
\end{equation}
and
\begin{equation}
(F_{y}^{\sigma\sigma'})_{mn}=(F_{y}^{\sigma'\sigma})_{nm}\,.
\end{equation}
Analog properties occur also for the matrix elements of the operators
$I[F_{\alpha}^{\sigma\sigma'}]$.  In particular, for delta impurities
we have in addition
\begin{equation}
(I[F_{\alpha}^{\sigma\sigma'}])_{mn}=(I[F_{\alpha}^{\sigma\sigma'}])_{nm}\,,
\end{equation}
which implies $I[F_{x}^{\sigma\sigma}]=0$.

\section{Some CNA results} \label{cnaresults}
The matrix elements of the electric modulation potential $V(x)=U \cos (K X_0)$
in the Landau representation are \cite{Gossmann98:1680}
\begin{equation}
V_{mn}(X_0) = U \, E_{mn}\Big( \frac{1}{2} \ell^2 K^2\Big) \, \cos \Big(KX_0
+(m-n) \frac{\pi}{2} \Big) \, ,
\end{equation}
with $E_{mn} (z)$ defined in Eq.~(\ref{enmvonz}). For weak modulation,
perturbation expansion yields to lowest order in $U$
\begin{equation}
E_{n,X_0}^{(1)} = \epsilon_n + U \, E_{nn}(\ell^2 K^2/2) \cos (KX_0) \, ,
\end{equation}
with $\epsilon_n = \hbar \omega_c (n+1/2)$. Since for $R_n \equiv \ell
\sqrt{2n+1} \gg 1/K$ one has the asymptotic expression \cite{Zhang90:12850}
$ E_{nn}(\ell^2 K^2/2) \approx \cos (KR_n -\pi/4) / (\pi K R_n /2)^{1/2}$, one
expects flat Landau bands, if the {\em flat-band condition} $2 R_n = a \,
(\lambda -1/4)$ holds for a $\lambda =1,\,2,\, \dots$. The states of the
modulated basis are,\cite{Zhang90:12850} to first order in $U$,
\begin{equation}
|n X_0) ^{(1)} = |n X_0\rangle + \sum_{n' \neq n} |n' X_0\rangle \,
\frac{V_{n'n}(X_0)}{\epsilon_n - \epsilon_{n'}} \, .
\end{equation}
In first order of $U$, this leads to a non-vanishing diagonal matrix element
of $v_y$ in the modulated basis,  given by $ (nX_0|v_y
|nX_0)= -(m \omega_c)^{-1} d E_{n,X_0}^{(1)} /dX_0$. For the off-diagonal
velocity matrix elements one obtains
\begin{equation}
|(mX_0|v_\alpha |nX_0)|^2  =| \langle mX_0|v_\alpha |nX_0 \rangle|^2 +
 O(U/\hbar \omega_c) \,  ,
\end{equation}
where the leading contributions $| \langle mX_0|v_\alpha |nX_0 \rangle|^2$
are independent of $U$ and identical for $v_x$ and $v_y$, see
 Eq.~(\ref{velomat}). Thus,  for weak modulation the scattering conductivities
 $\sigma^{scat}_{xx}$ and $\sigma^{scat}_{yy}$ are approximately the same. For
 stronger modulation this is no longer true, even within the CNA.

Within the CNA the imaginary part of Eq.~(\ref{scba}) reads
\begin{equation} \label{cnascba}
{\rm Im}\, \Sigma^-(E)= \Gamma_0^2 \sum_n \int_0^a \!
\frac{dX_0}{a}\, {\rm Im}\, G_{nn}^- (X_0,E) \, ,
\end{equation}
and the DOS is given by 
$D(E)= D_0\,{\rm Im} \Sigma^-(E)/(\pi\Gamma_0^0)$.
For an estimate of the scattering and band conductivities we assume that the
collision broadening ${\rm Im} \Sigma^-(E)$ of the spectral functions $ {\rm
  Im} G_{nn}^- (X_0,E) /\pi$ is much smaller than the modulation-induced
bandwidth ($\sim  U \,| E_{nn}(\ell^2 K^2/2)|$), and that the latter is much
smaller than $\hbar \omega_c$. Then the dominant contribution to
Eq.~(\ref{condcna}) comes from the Landau bands at the energy $E$. Let us
assume that $E$ lies in the band $n_F$, so that $|E-E_{n_F,X_0} -{\rm
  Re}\Sigma(E)| \ll \hbar \omega_c$. Then, the leading order contributions to
the scattering conductivity are obtained for $n=n_F$ and $n'=n_F \pm 1$ (and
$n$ and $n'$ interchanged), and we may approximate
\begin{equation}
 {\rm Im} G_{n'n'}^- (X_0,E) \approx {\rm Im} \Sigma^-(E) /(\hbar \omega_c)^2
 \,, 
\end{equation}
Inserting this and the off-diagonal velocity matrix elements from
Eq.~(\ref{velomat}), we obtain
\begin{equation} \label{siscatt}
\sigma_{\alpha \alpha}^{scat}(E) \approx \frac{e^2}{h} \, (2 n_F +1)
\frac{\Gamma_0^0}{\hbar \omega_c} \left[\frac{D(E)}{D_0} \right]^2 \, .
\end{equation}
This is a general result and holds for unmodulated systems \cite{Ando82:437}
and for a 2DEG with a two-dimensional superlattice as
well.\cite{Gerhardts91:5192} It shows that the scattering conductivity is
proportional to the {\em scattering rate} $1/\tau_{sc} =\Gamma_0^0/2 \hbar$,
and that its maxima and minima follow those of the DOS.

The band conductivity $\sigma_{yy}^{band}$ is dominated by the term containing
 the square of the spectral function ${\rm Im}\, G_{n_F n_F}^- (X_0,E)/\pi$.
 Considered as  a function  of $X_0$,
the latter has  $\delta$-function-like peaks at the values $X_F$ satisfying
\begin{equation} \label{xsubf}
 E-E_{n_F,X_F} -{\rm Re}\Sigma (E) =0 \, .
\end{equation}
To obtain an estimate (actually an upper limit) of
 $\sigma_{yy}^{band}$, we take the square of the velocity matrix elements and
one of the ${\rm Im}\, G_{n_F n_F}^- (X_0,E)$ factors at $X_0=X_F$, i.e., we
 approximate 
 ${\rm Im}\, G_{n n}^- (X_0,E) \approx {\rm Im}\, G_{n_F n_F}^- (X_F,E) =1/
 {\rm Im}\Sigma^-(E)$. Then we obtain with Eqs.~(\ref{condcna}) and
(\ref{driftve}) 
\begin{equation} \label{siband1}
\sigma_{yy}^{band}(E) \approx \frac{ e^2}{h}\,
 \left| \frac{2\ell}{\Gamma_0}\, \frac{dE_{n_F,X_0}}{dX_0} \right|^2_{X_0=X_F} \, ,
\end{equation}
which according to Eq.~(\ref{xsubf}) depends on $E$ (at $T \rightarrow 0$ on
$E_F$,) via $X_F =X_F(E)$. This shows already, that $\sigma_{yy}^{band}(E)$
is proportional to the {\em scattering time} $2\hbar / \Gamma_0^0$, and that it
becomes small for flat bands and near the band edges, where the energy
dispersion becomes flat. This is opposite to the DOS, which becomes large at
flat bands and near the band edges (VHS).
 
If Eq.~(\ref{xsubf}) holds for $E$ well within a Landau band, i.e.,
sufficiently far from the band edges, we may approximate ${\rm Im}\, G_{n_F
  n_F}^- (X_0,E)/\pi \approx \delta(X_0-X_F)/ |dE_{n,X_0}/dX_0|_{X_0=X_F} $
to obtain
\begin{equation}
D(E) \approx 2/[\pi a \ell^2 \,| dE_{n_F,X_0} / dX_0|_{X_0=X_F}] \, ,
\end{equation}
where we assumed that  Eq.~(\ref{xsubf}) is satisfied  at $X_0 = \pm X_F$.
Inserting this into Eq.~(\ref{siband1}) yields
\begin{equation} \label{siband2}
\sigma_{yy}^{band}(E) \sim \frac{e^2}{h} \,  \frac{\hbar
  \omega_c}{\Gamma_0^0} \left[ \frac{4 \ell\,  D_0}{a D(E)} \right]^2 \, ,
\end{equation}
which is a reasonable estimate at energies well inside the 
modulation-induced Landau bands.

If we repeat the analysis of the band conductivity for
overlapping Landau bands, i. e. if we drop the condition for the bandwidth
$U \,| E_{nn}(\ell^2 K^2/2)| \ll \hbar \omega_c$ we arrive at
\begin{equation} \label{overlap}
\sigma_{yy}^{band}(E) \approx \frac{ e^2}{h}\,
\left( \frac{2\ell}{\Gamma_0} \right)^2 \, \frac{\left| \sum_n
\frac{dE_{n,X_0}}{dX_0} \right|_{E_{n,X_0}=E}}{\left| \sum_{n'}
\left(\frac{dE_{n',X_0}}{dX_0} \right)^{-1} \right|_{E_{n',X_0}=E}} \; ,
\end{equation}
The difference of this expression to Eq.~(\ref{siband1}) is the contribution of
more than one Landau band at the same time to the total density of states
(denominator) and of bands with differing group velocities to the conductivity
(numerator).


%
\begin{figure}
\caption{Diagrams for the generalized SCBA: Dyson 
[(\protect\ref{greenf}) and (\protect\ref{scba})] and Bethe-Salpeter 
[(\protect\ref{vertexf}) and (\protect\ref{vertexc})] 
equations.}
\end{figure}
\begin{figure}
\caption{(a) Density of states at the Fermi level for the electric
modulation, $D_0=m/(\pi\hbar^2)$.
(b) The energy spectrum for $B_0=0.81$ T. The dashed
horizontal line shows the Fermi level.}
\end{figure}
\begin{figure}
\caption{Conductivities for the electric modulation.}
\end{figure}
\begin{figure}
\caption{Resistivities for the electric modulation.}
\end{figure}
\begin{figure}
\caption{Resistivities for the magnetic modulation in the regime
of high band conductivity.}
\end{figure}
\begin{figure}
\caption{(a) Energy spectra for the magnetic modulation with GaAs 
parameters, corresponding to Fig.\ 5. (b) The density of states
at the Fermi level for $r_0=10$ nm.}
\end{figure}
\begin{figure}
\caption{Resistivities for the magnetic modulation in the regime of
low band conductivity.}
\end{figure}
\begin{figure}
\caption{An energy spectrum for the magnetic modulation with InAs 
parameters, corresponding to Fig.\ 7, with $B_0=3$ T.}
\end{figure}
\begin{center}
\newpage
\epsffile{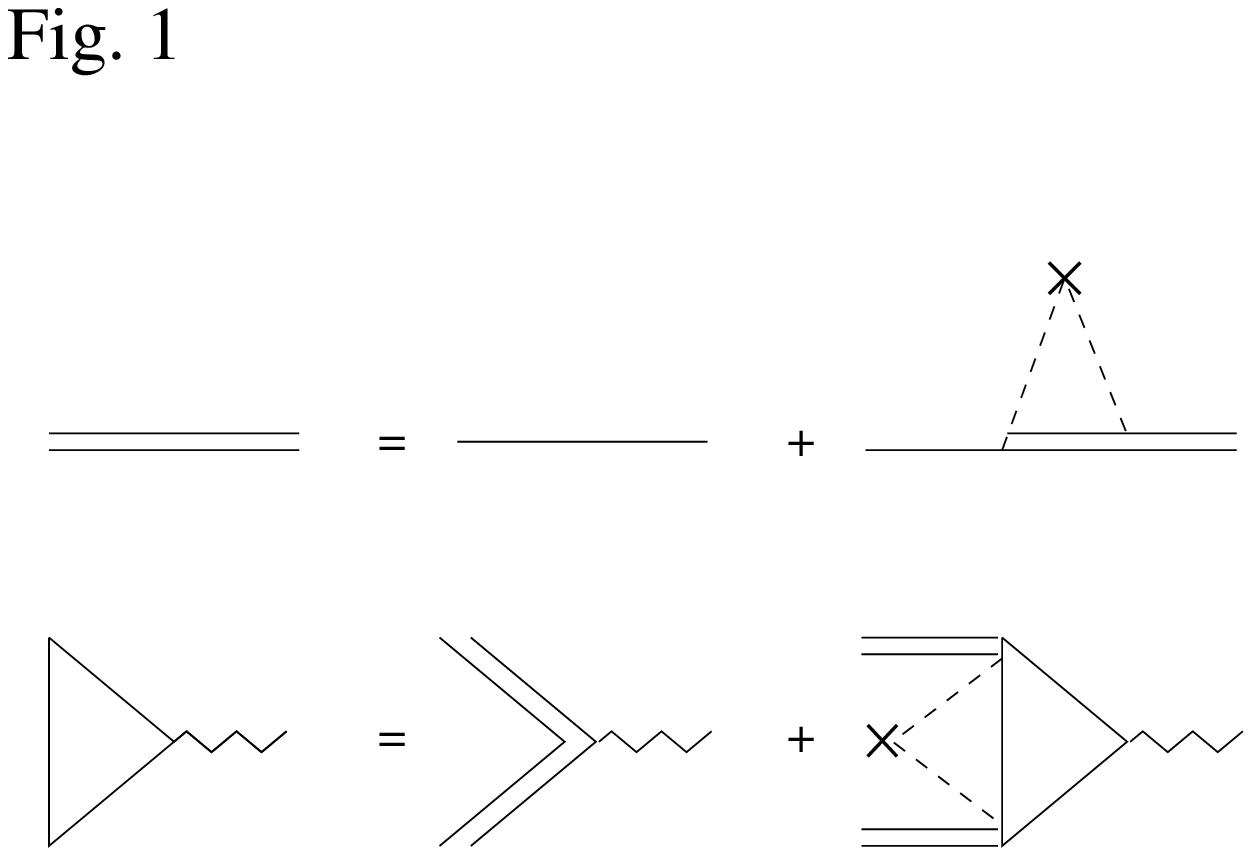}
\newpage
\epsffile{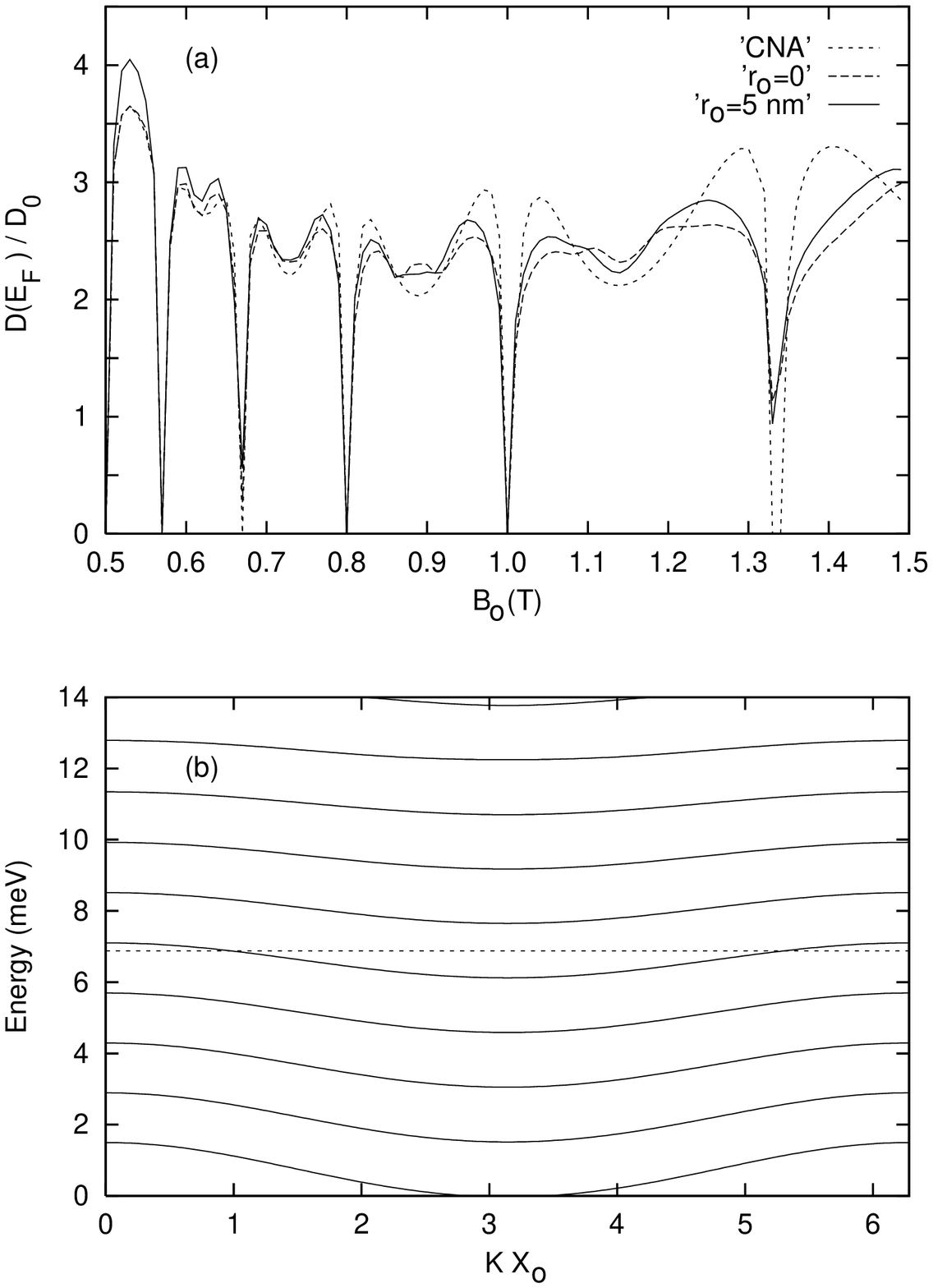}
\newpage
\epsffile{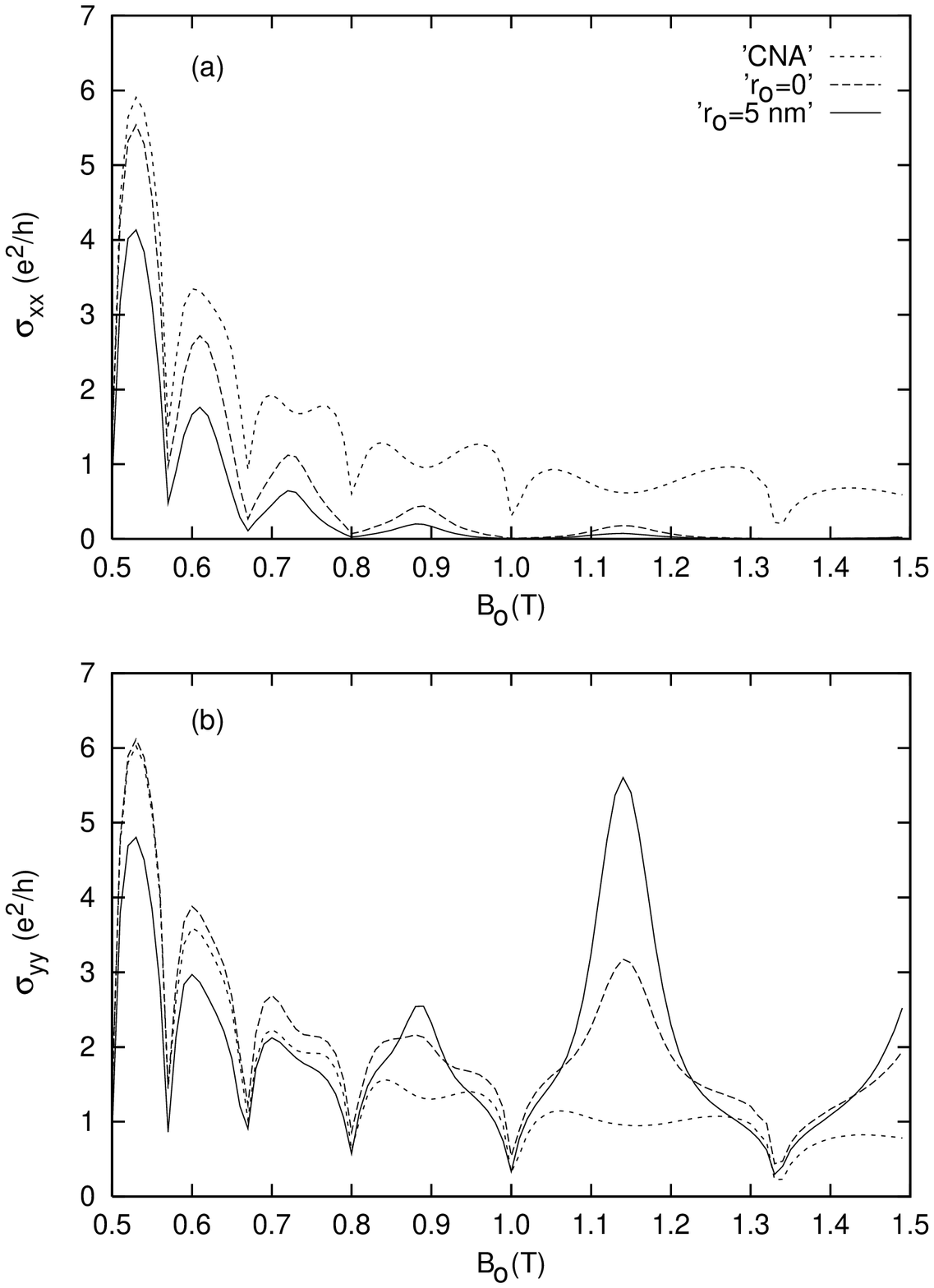}
\newpage
\epsffile{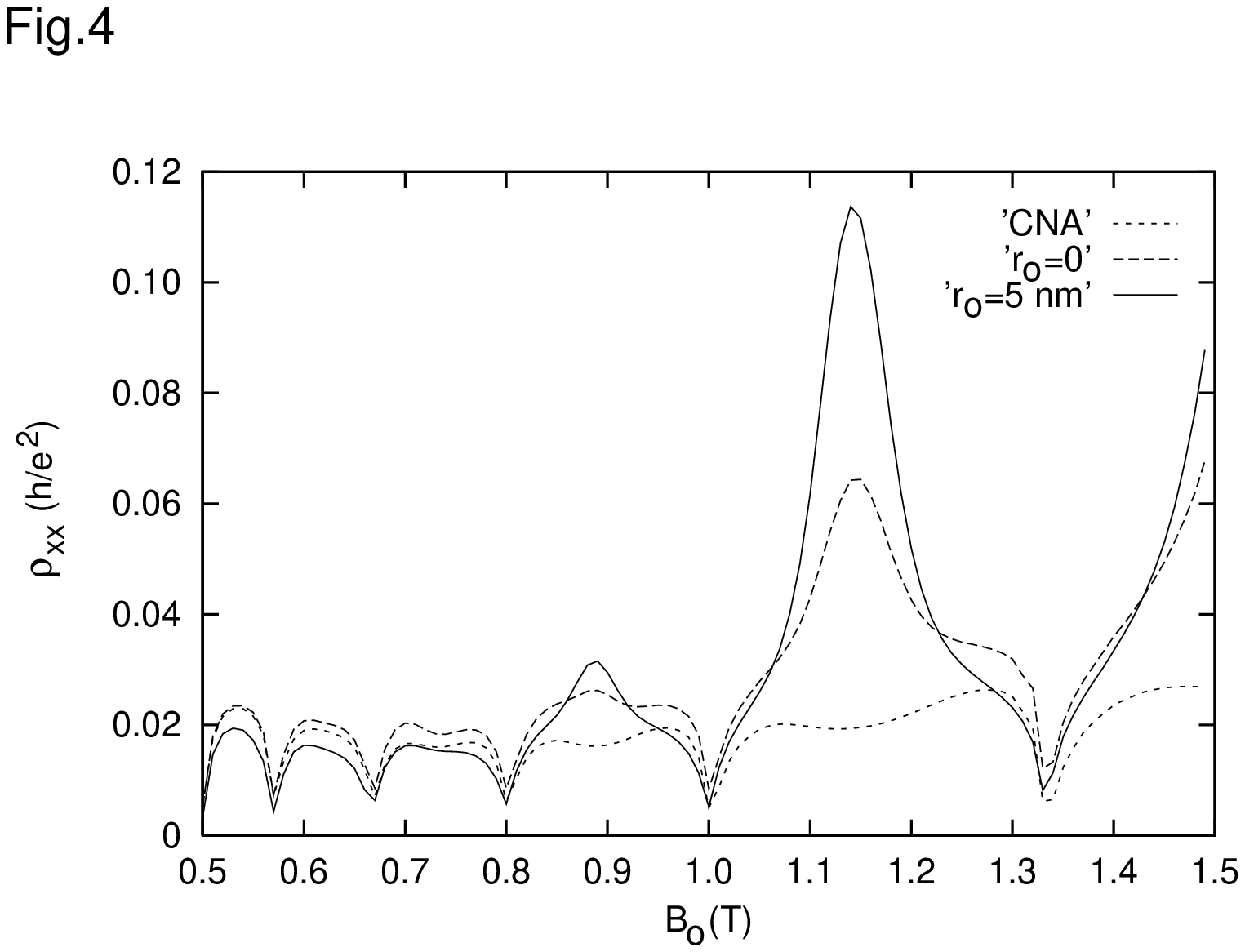}
\newpage
\epsffile{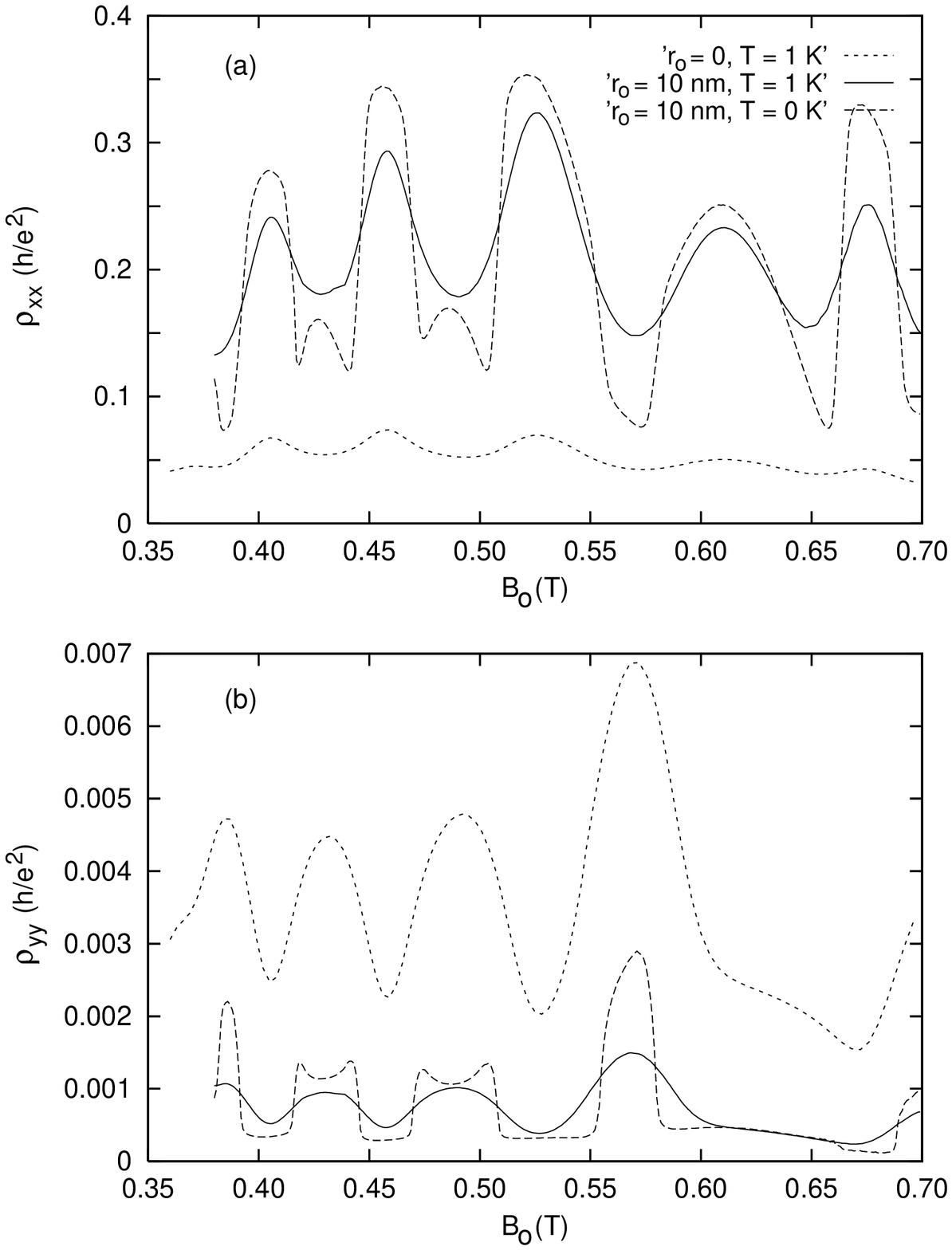}
\newpage
\epsffile{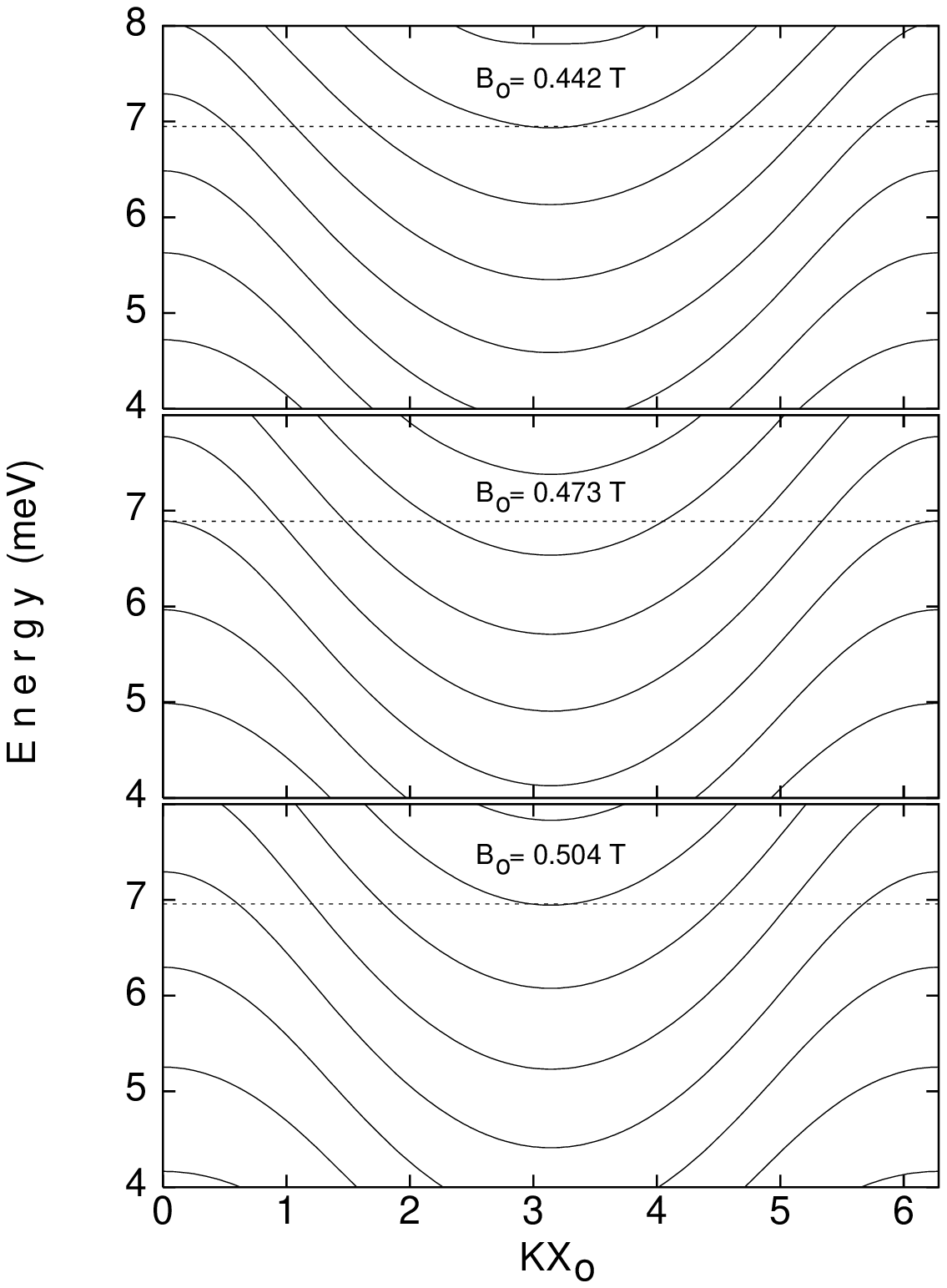}
\newpage
\epsffile{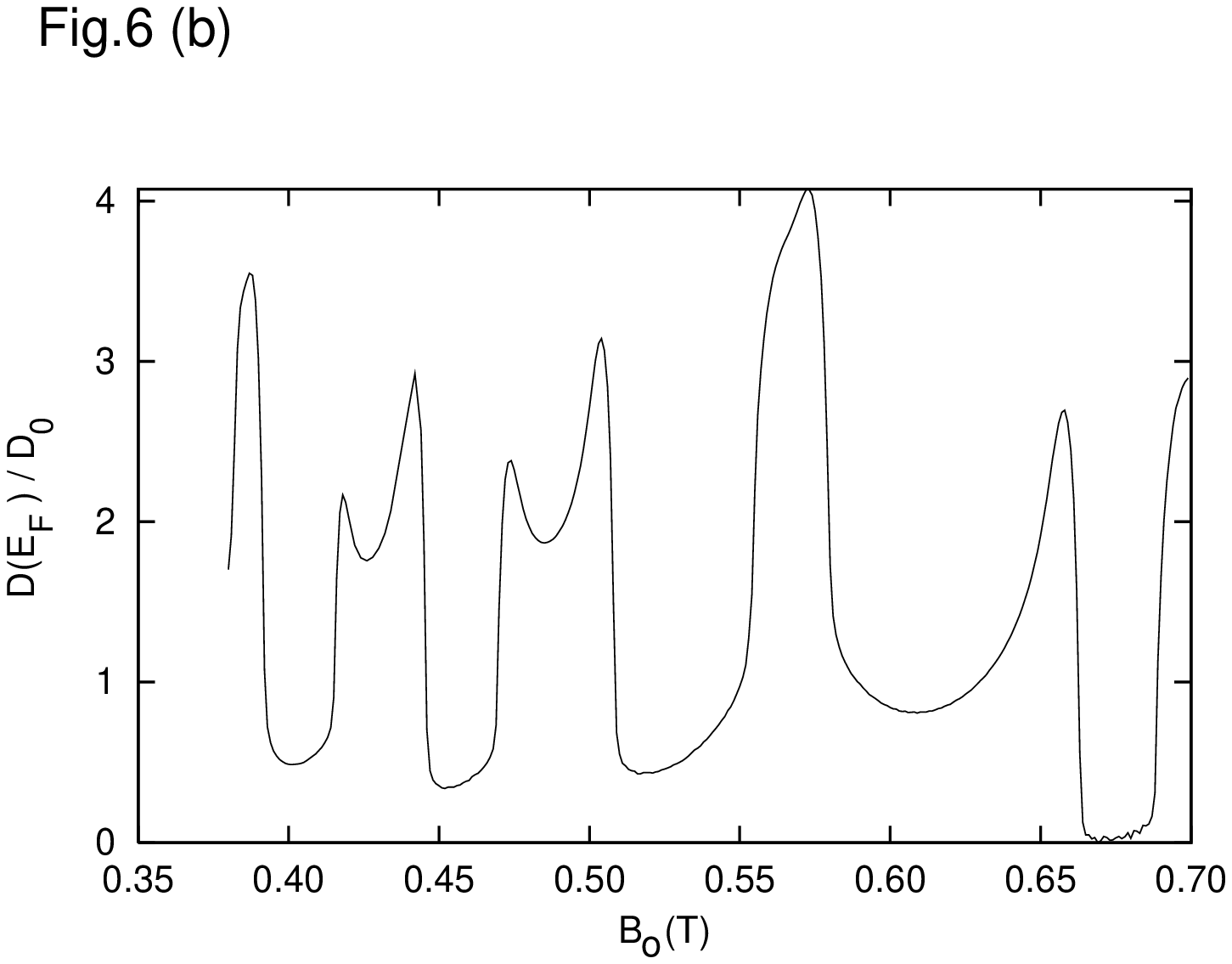}
\newpage
\epsffile{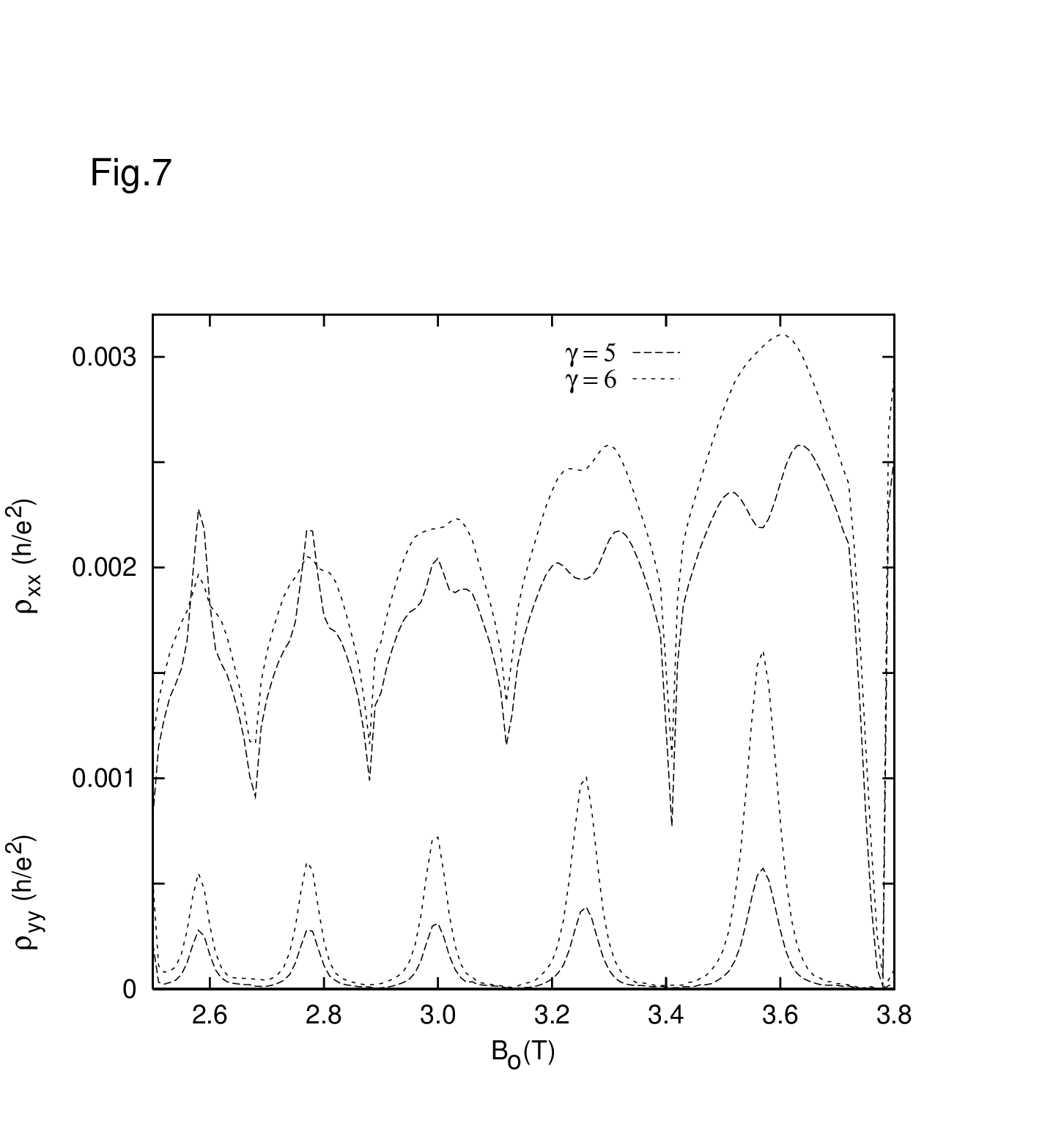}
\newpage
\epsffile{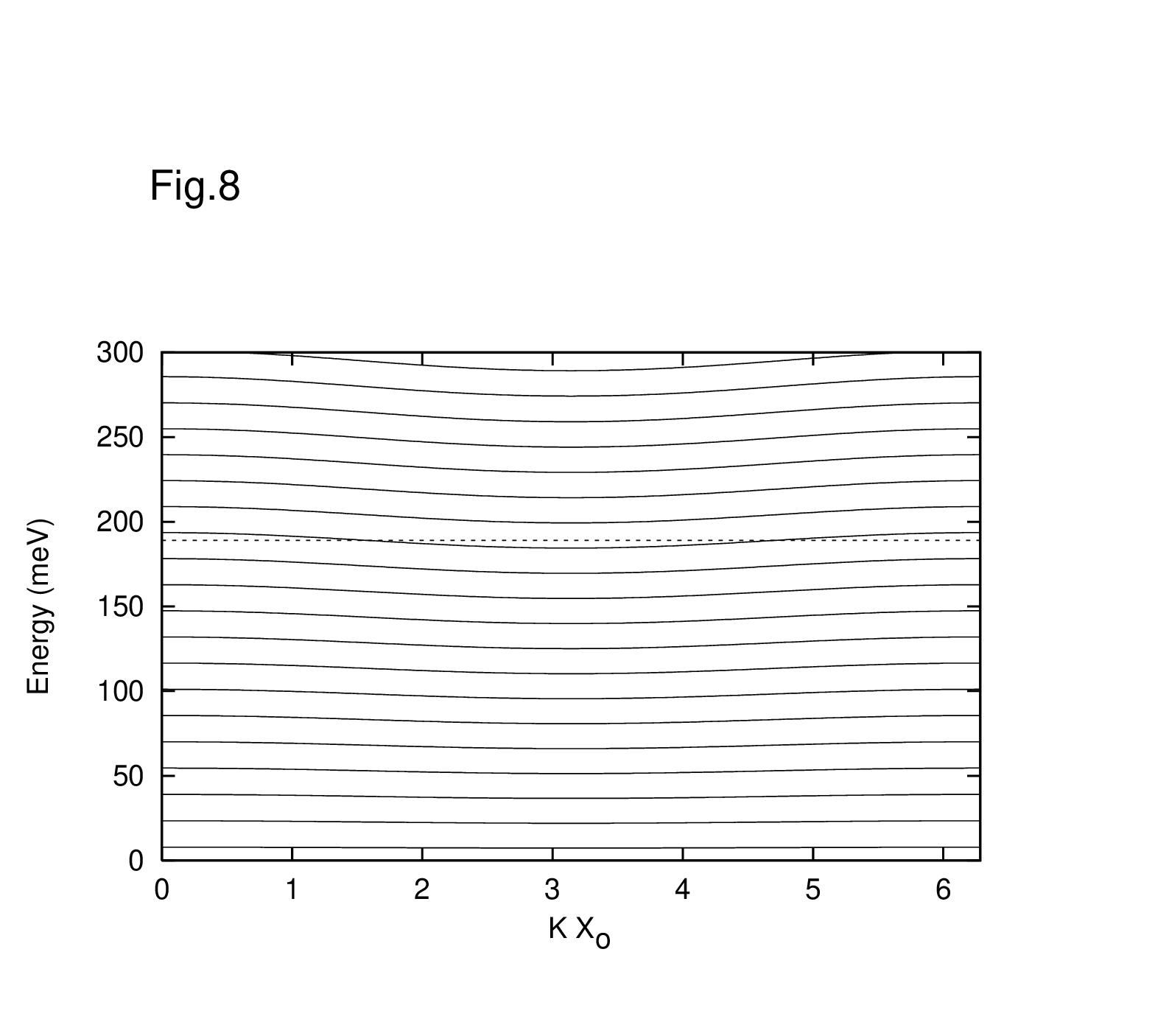}
\end{center}

\end{document}